\documentclass[referee]{aa}  

\usepackage{graphicx}
\usepackage{txfonts}
\begin{document}

\title{Minimum Orbital Intersection Distance: Asymptotic Approach}
\titlerunning{MOID: Asymptotic Approach}
\authorrunning{J.~M. Hedo \& E. Fantino et al.}

\author{J.~M. Hedo\inst{1}
        \and
        E. Fantino\inst{2}
				\and
				M. Ru\'iz\inst{1}
								\and
        J. Pel\'aez\inst{1}}

   \institute{Space Dynamics Group, ETSIAE, Technical University of Madrid, Pza. Cardenal Cisneros 3, 28040 Madrid, Spain\\
	 \email{josemanuel.hedo@upm.es}
        \and
         Aerospace Engineering Department, Khalifa University of Science and Technology, P.O. Box 127788, Abu Dhabi, United Arab Emirates\\
				\email{elena.fantino@ku.ac.ae}}

   \date{Received ---; accepted ---}

 
  \abstract{
  The minimum orbital intersection distance is used as a measure to assess potential close approaches and collision risks between astronomical objects. Methods to calculate this quantity have been proposed in several previous publications. 
The most frequent case is that in which both objects have elliptical osculating orbits. When at least one of the two orbits has low eccentricity, the latter can be used as a small parameter in an asymptotic power series expansion. The resulting approximation can be exploited to speed up the computation with negligible cost in terms of accuracy. This contribution introduces two asymptotic procedures into the SDG-MOID method presented in a previous article, it discusses the results of performance tests and their comparisons with previous findings. The best approximate procedure yields a reduction of 40\% in computing speed without degrading the accuracy of the determinations. This remarkable result suggests that large benefits can be obtained in applications involving massive distance computations, such as in the analysis of large databases, in Monte Carlo simulations for impact risk assessment or in the long-time monitoring of the MOID between two objects.}

   {}


   \maketitle
%

\section{Introduction}
The minimum orbital intersection distance (MOID) at a given epoch between two objects orbiting a common primary is the distance between the closest points of their osculating orbits.
The problem of computing the MOID is as old as Kepler's laws. However, the need for a quick and accurate solution has increased as its role in several branches of Celestial Mechanics and Astrodynamics has become more and more prominent.
Nowadays, the MOID is used primarily in the following two contexts:
\begin{itemize}
  \item To provisionally discard objects from large space debris catalogs when collisions with a given target can be excluded \citep{hoots:1984, Woodburn:2009, Casanova:2014}. According to \citet{Gorno:2019}, ``the ever growing population of space debris (or space junk), especially in the LEO zone, is posing serious issues to the survival of man-made objects in space. Every mission planner has to take into account the possibility of the satellite to collide with relatively small pieces of debris which, traveling at orbital speeds, could jeopardize the survival of the mission''. The Center for Space Standards \& Innovation (CSSI) offers the Satellite Orbital Conjunction
Reports Assessing Threatening Encounters in Space (SOCRATES) service \citep{Kelso:2006}, intended to help satellite operators avoid undesired close approaches through advanced mission planning. According to \citet{Kelso:2006}, ``hundreds of conjunctions within 1~km occur every week, with little or no intervention, putting billions of dollars of space hardware at risk, along with their associated missions''.
  \item To predict possible close encounters of asteroids and comets with the planets (mainly, Earth and Mars for the asteroids and Jupiter for the comets) prior to the study of their dynamics \citep{Milani:1989, Harris:1998, Milani:2000, Wlodarczyk:2013}. The MOID between the asteroid and Earth turns out to be one of the most important parameters when assessing the impact risk \citep{Sitarski}.
\end{itemize}
The calculation of the MOID involves two Keplerian orbits, represented by their classical orbital elements. 
The orbital elements change over time as a result of perturbations. In the case of the asteroids, for example, \citet{Sekhar:2016} show that Einstein's General Theory of Relativity predicts a significant orbital precession if the object has both a moderately low perihelion and a small semimajor axis, and observes that such effect ``could play an important role in the calculations pertaining to MOID and close encounter scenarios in the case of certain small solar system bodies (depending on their initial orbital elements)''. Due to the time evolution of the orbital elements, the MOID itself is a function of time. Thus, understanding the evolution of the MOID over long time frames is important for impact risk assessment. In the case of asteroids, such MOID monitoring allows to better assess the risk of a possible impact with Earth. Recently, Monte Carlo methods have been implemented to calculate the impact probability of an asteroid with  Earth \citep{Rickman:2014}. 

The orbit of an asteroid is known with an uncertainty that depends on the quality of the observations and that reflects in the uncertainty of the MOID. For its assessment, the covariance matrix must be propagated, with the inherent requirement to explicitly compute the derivatives of the MOID with respect to some variables. \citet{Bonanno:2000} uses an \textit{approximate analytical} expression to obtain the required partial derivatives of the MOID. Alternatively, the differentiation can be carried out numerically, in which case it is  beneficial to employ fast, accurate and safe algorithms to calculate the MOID without increasing the uncertainty brought by the observational data. With this respect, it is worth remembering that a singularity appears in the computation of the uncertainty of the MOID in the orbit-crossing case. This problem has been studied and solved by \citet{Gronchi:2007}.

Using models currently accepted for the long-term dynamical evolution of the solar system, the MOID computation allows to determine the cometary impact rates on the Moon and on the terrestrial planets during the late heavy bombardment \citep{Rickman:2017}. Fast MOID computation is in high consideration and demand also to facilitate the analysis of large catalogs of objects. 
Obviously, accuracy is also a key issue. Here, the latter is meant to encompass both precision (low random errors) and trueness (low systematic errors) \citep{ISO5725}.

In summary, a fast and accurate MOID computation method is an asset in several common problems. Many methods for computing the MOID have been published in the literature over the past seven decades. The majority are approximate numerical methods \citep{Sitarski, Dybczinski}. However, more recently, algebraic \citep{Kholshevnikov:1999, Gronchi:2005} and analytical \citep{Bonanno:2000, Armellin:2010} approaches have been established, and some hybrid methods have appeared \citep[see, e.g.,][]{Derevyanka:2014}.

The Space Dynamics Group (SDG) of the Technical University of Madrid developed a fast and accurate numerical method for the computation of the MOID. This method is called SDG-MOID \citep{Hedo:2018} and is based on two algorithms: 
\begin{description}
  \item[Algorithm 1] determines the distance between a point and an ellipse in three-dimensional space by composing two orthogonal distances:
      \begin{enumerate}
       \item the distance from the point to its orthogonal projection onto the plane of the ellipse;
       \item the distance from the ellipse to the projected point.
       \end{enumerate}
The former is trivial to determine, whereas the latter (which for the sake of brevity will be referred to as the in-plane distance)
can be obtained in closed form with elementary functions only under specific circumstances\footnote{These circumstances (which are very uncommon) are the case $e=0$ (circular orbit) and that in which the projected point lies on a symmetry axis.}, the general scenario requiring application of an iterative method to the solution (as accurate as desired) of a transcendental equation. 

  \item[Algorithm 2] calculates the minimum distance between two confocal ellipses by:
  \begin{enumerate}
    \item discretizing one of them in a finite set of points by means of a uniform partition of the interval $[0,2\pi]$ where its eccentric anomaly is defined --see Eqs.~\ref{eq:2-2-1} and \ref{eq:2-2-2} below-- and computing the corresponding distances to the other ellipse by means of Algorithm 1: this yields a discrete function approximation of the distance;
    \item identifying the partition intervals containing local minima of the distance;
    \item minimizing the distance function in these intervals using the golden-section search;
    \item selecting the minimum minimorum.
  \end{enumerate}
\end{description}
This work considers asymptotic expansions for the computation of the in-plane distance component in Algorithm 1 of the SDG-MOID method. These asymptotic procedures are tested to assess the gain in computing speed, the corresponding accuracy loss, and hence the benefits of their introduction: they are worth to be adopted if a positive balance between advantages and disadvantages is achieved.
Our tests employ the NEODyS (Near Earth Object - Dynamics Site) catalog \citep{NEODyS} for MJD 58600 to compute the position uncertainty and the Earth MOID of NEOs.
Improvements of Algorithm 2 are also being undertaken: they aim at preserving the accuracy by decreasing the resolution of the discretization and by limiting it to the vicinity of the minimum minimorum. However, these enhancements are under development and out of the scope of this contribution.

The paper is structured as follows. Section~\ref{sec:MF} illustrates the mathematical foundations of the two asymptotic expansions here proposed as approximate procedures to compute the in-plane distance component. These methods constitute one-parameter families of procedures, the parameter being the number of terms retained in the expansions. Section~\ref{sec:accuracy} assesses the minimum MOID accuracy (trueness) needed in order not to deteriorate the total uncertainty of the starting orbital data. Section~\ref{sec:tests} presents the results of the accuracy and computing speed tests on the two families of asymptotic expansions and compares them with the performance of the exact in-plane distance determination carried out with SDG-MOID. The family exhibiting the best performance is tested for accuracy and computing speed on the Earth MOID of NEOs and the results are compared with the SDG-MOID determinations. Section~\ref{sec:app} illustrates the  benefits that the asymptotic expansions can bring when applied to the time monitoring of the MOID through the presentation of an application to the space debris problem. 
The paper closes (Sect.~\ref{sec:conclu}) with a summary and a discussion of the relevance of the results.

\section{Mathematical foundations}
\label{sec:MF}
This section summarizes the theory of asymptotic expansions in powers of a small parameter and describes in detail the application of two such methods to the determination of the in-plane distance component.

\subsection{Asymptotic expansions in powers of a small parameter}
Given the function $f : \mathbb{R}^{n+1} \mapsto \mathbb{R}$ and
\begin{equation}
    z=f(x;y_1,y_2,\dots, y_n),
\end{equation}
with $x \in \mathbb{R}: x\ll 1$ the independent variable, $z\in\mathbb{R}$ the dependent variable and $y_k\in\mathbb{R}\ (k=1,2,\dots,n)$ a set of parameters, the asymptotic expansion of $z$ in powers of $x$ is defined as:
\begin{equation}
    z \sim  \sum_{i=0}^{N} c_i(y_1,y_2, \dots, y_n)\, x^i,
\end{equation}
where $N$ is the order of the expansion, i.e., the highest power of $x$ among the terms that appear in the expansion, and $c_i\sim O(1)\ (i=0,1,2,\dots,N)$ are the functional coefficients.
$N$ is used to control the accuracy of the asymptotic expansion. The advantages of the truncation of the expanded function are:
\begin{itemize}
  \item a higher speed of the computation of the approximate function compared with the original function;
  \item a more treatable mathematical expression for algebraic manipulation and computation.
\end{itemize}
The main drawback consists in the introduction of truncation errors, i.e., the expansion only provides an approximation to true values. The smaller the value of $x$ and the higher the value of $N$, the better the approximation of $z$ obtained from the corresponding truncation.
The procedure is useful if the improvement in computing speed compensates for the inherent loss of accuracy.

Several choices for the asymptotic expansion exist. In this paper, we consider the two most suitable options, i.e., an expansion in the critical eccentric anomaly and an expansion in the minimum distance. They are described here below.

\subsection{Expansion in the critical eccentric anomaly}
\label{ss:2.2}
Let $\mathfrak{e}$ be an ellipse in the standard Cartesian parametric representation relative to its symmetry axes:
\begin{eqnarray}
  x(u) & = &  a \cos(u),               \label{eq:2-2-1}\\
  y(u) & = & \sqrt{1-e^2}\, a \sin(u), \label{eq:2-2-2}
\end{eqnarray}
where $u$ is the eccentric anomaly, $a$ is the semi-major axis and $e\ll1$ is the eccentricity. Let $P$ be a coplanar point with Cartesian coordinates $(\alpha,\beta)$ in the previous axes. Find the eccentric anomaly $u_*$ that minimizes the distance $\textrm{d}(P,E) : E \in\mathfrak{e}$ defined as
\begin{equation}
    \textrm{d}(u; a,e,\alpha,\beta) = \sqrt{(\alpha-x(u))^2+(\beta-y(u))^2}.
    \label{eq:2-2-3}
\end{equation}
This problem was addressed as part of the presentation of the SDG-MOID method \citep{Hedo:2018}. The main conclusions of that study can be summarized as follows:
\begin{enumerate}
\item $\textrm{d}(u)$ is $2\pi$-periodic and, therefore, it has an even number of extreme points. All global maxima and minima are always critical points.
\item The critical points $u^*$ of Eq.~\ref{eq:2-2-3} (equivalent to the critical points of its square) must satisfy the transcendental equation
    \begin{equation}
    f(u;a,e,\alpha,\beta) =  a\alpha \sin \left( u \right) - \sqrt {1-{e}^{2}}\,a\beta\cos \left(u \right)
                           - {e}^{2}{a}^{2}\sin \left(u\right)\cos\left(u\right) = 0.
		\label{eq:2-2-4}
    \end{equation}
Equation~\ref{eq:2-2-4} admits a minimum of two and a maximum of four solutions in the interval $[0,2\pi)$ (including, in the latter case, a possible double root resulting from the coincidence between one maximum and one minimum). Solving Eq.~\ref{eq:2-2-4} is equivalent to solving a polynomial equation of degree four. The latter admits exact solutions, but, as pointed out in \citet{Hedo:2018}, the loss of precision due to the rounding errors may change their character, e.g., from real
to complex. Since the interest is in real solutions, this approach cannot be employed.
\item The minima must also satisfy the condition
    \begin{equation}
        f'(u)>0.
				\label{ineq:2-2-5}
    \end{equation}
The geometric study showed that the absolute minimum (and the absolute maximum) always exists. Let $\mathcal{M}=\{u_i, i=1,2,\dots,m\}$ be the set of eccentric anomalies corresponding to the minima (they satisfy equations~\ref{eq:2-2-4} and \ref{ineq:2-2-5}). Then,
\item the eccentric anomaly $u_*$ yielding the minimum distance in $\mathcal{M}$ is
    \begin{equation}
    u_* =  \{ u_i \in \mathcal{M} : d(u_i)\le d(u_j), \forall j\not= i \}.
    \end{equation}
    The symbol $\le$ takes into account the case in which multiple $u_*$'s are associated with the same minimum distance.
\end{enumerate}
The results can be summarized as follows:
\begin{itemize}
  \item If $e=0$ (circular orbit), the solution is simply $u_*=\arctan(\beta/\alpha)$.
  \item If $e>0$ and $\alpha=0$ or $\beta=0$ ($P$ on a symmetry axis of $\mathfrak{e}$), closed-form solutions for $u_*$ always exist.
  \item If $e>0$, $\alpha\not=0$ and $\beta\not=0$, numerical techniques constitute the only method to solve the equation, with the additional requirement that $u_*$ be in the same quadrant as $P$. In particular, iterative methods offer the best results (i.e., the fastest and the most accurate), and a suitable seed for such numerical approximations is $u_0=\arctan(\beta/\alpha)$.
\end{itemize}
Equation~\ref{eq:2-2-4} defines an implicit function $u_*(a,e,\alpha,\beta)$ and, when ${e\ll 1}$, a truncated series expansion in powers of $e$ is convenient:
\begin{equation}
  u_*^{(N)}(a,e,\alpha,\beta) = \sum_{i=0}^{N} c_i(a,\alpha,\beta)\, e^i.
\label{eq:2-2-7}
\end{equation}
The parameter $N\in\mathbb{N}$ controls the finite number of terms in the truncated series, whereas the $c_i(a,\alpha,\beta)$ coefficients must be computed to approximate the critical eccentric anomaly.

Substituting $u_*$ as given by Eq.~\ref{eq:2-2-7} into Eq.~\ref{eq:2-2-4} yields the expansion in powers of $e$. Since $f=\sum_{i=0}^N f_i e^i$ must satisfy Eq.~\ref{eq:2-2-4}, the $c_i$'s can be calculated by successive cancellation of the $f_i$'s in increasing order.
This problem has been solved for $N$ up to 6 using the {\sc Maple} 2017.3 math software. The result is:
\begin{eqnarray}
  c_i & = & 0 \ \text{if }i \text{ is odd}, \label{eq:2-2-8} \\
  c_0 & = & \arctan\left(\frac{\beta}{\alpha}\right) \text{\ (case $e=0$ above)}, \\
  c_2 & = & \alpha\,\beta\, \left(
  \frac{a}{\left({\alpha}^{2}+{\beta}^{2}\right)^{3/2}} - \frac{1}{2\left({\alpha}^{2}+{\beta}^{2}\right)}  \right) \\
  c_4 & = & - \frac{\alpha\,\beta\, \left( -8\,{a}^{2}{\alpha}^{2}+ 8\,{a}^{2}{\beta}^{2}+{\alpha}^{4}+4\,{\alpha}^{2}
  {\beta}^{2}+3\,{\beta}^{4} \right) } {8 \left( {\alpha}^{2}+{\beta}^{2}  \right) ^{3}} + \frac {a\, \alpha\,\beta\,
  \left( {\beta}^{2}-\frac{1}{2}\alpha^2 \right)}
    { \left( {\alpha}^{2}+{\beta}^{2}  \right) ^{5/2}}, \\
  c_6 & = & \frac{\alpha\beta}{\left(\alpha^2+\beta^2\right)^{4}}
  \left[
  -\frac{\alpha^6}{16}
  - \alpha^4 \left( \frac{a^2}{2} +\frac {13 \beta^2}{48} \right) +
  {\alpha}^{2} \left( 4\,{a}^{2}{\beta}^{2} - \frac{25{\beta}^{4}}{48} \right)
  - \frac{3}{2}\,{a}^{2}{\beta}^{4} -\frac {5{\beta}^{6}}{16} \right] +  \nonumber \\
	& + & \frac{a\,\alpha\,\beta}{\left( \alpha^2+ \beta^2 \right)^{9/2}}
  \left[ -\frac{\alpha^6}{8} + \alpha^4\left(a^2 - \frac{9\beta^2}{8} \right) 
  - \frac{23{a}^{2}{\alpha}^{2}{\beta}^{2}}{6} + a^2 \beta^4+ \beta^6
  \right]. \label{eq:2-2-12}
 \end{eqnarray}
After substituting Eqs.~\ref{eq:2-2-8} to \ref{eq:2-2-12} into Eq.~\ref{eq:2-2-7}, the Euclidean distance function (Eq.~\ref{eq:2-2-3}) with $u=u_*^{(N)}$ yields the following parametric family of procedures for the approximate computation of the minimum distance:
\begin{equation}
\textrm{d}_*^{(N)}(a,e,\alpha,\beta) = \textrm{d}(u_*^{(N)}(a,e,\alpha,\beta); a,e,\alpha,\beta),
\label{eq:2-2-13}
\end{equation}
in which $N=0,2,4,6$ is the parameter. The selection of $N$ is conducted as a trade-off between efficiency and accuracy: the lower its value, the more mathematically treatable the final expression contained in Eq.~\ref{eq:2-2-3}, but the lower the accuracy of the resulting distance approximation.

\subsection{Minimum distance expansion}
Following \citep{Hedo:2018}, an alternative approach to the computation of the critical points of the distance function consists in solving the following system of two algebraic equations:
\begin{eqnarray}
  \dfrac{x^2}{a^2} + \dfrac{y^2}{a^2(1-e^2)} & = & 1, \label{eq:2-3-1} \\
  a^2 e^2 xy -\alpha\,a^2\,y + \beta\,a^2(1-e^2)\, x &= & 0. \label{eq:2-3-2}
\end{eqnarray}
If $e=0$, the solution corresponding to the minimum distance is trivial:
\begin{eqnarray}
    x_0(a,\alpha,\beta) & = &\frac{a\alpha}{\sqrt{\alpha^2+\beta^2}}, \label{eq:2-3-3} \\
    y_0(a,\alpha,\beta) & = &\frac{a \beta}{\sqrt{\alpha^2+\beta^2}}. \label{eq:2-3-4}
\end{eqnarray}
If $0 < e \ll 1$, let us consider the following decomposition:
\begin{eqnarray}
  x_* & = & x_0 + \xi \ :\ \xi  \sim O(e), \label{eq:2-3-5} \\
  y_* & = & y_0 + \eta\ :\ \eta \sim O(e). \label{eq:2-3-6}
\end{eqnarray}
Substituting Eqs.~\ref{eq:2-3-5} and \ref{eq:2-3-6} with Eqs.~\ref{eq:2-3-3} and \ref{eq:2-3-4} into Eqs.~\ref{eq:2-3-1} and \ref{eq:2-3-2} leads to a system of two equations to be solved with respect to $\xi$ and $\eta$:
\begin{alignat}{2}
  \xi &= \xi(a,e,\alpha,\beta), \label{eq:2-3-7}\\
  \eta &= \eta(a,e,\alpha,\beta). \label{eq:2-3-8}
\end{alignat}
Hence, the square of the distance function (Eq.~\ref{eq:2-2-3}) can be written as:
\begin{equation}
  \textrm{d}^2 = (\alpha - x_0 - \xi)^2+(\beta - y_0 - \eta)^2 = \textrm{d}^2(a,e,\alpha,\beta).
  \label{eq:2-3-9}
\end{equation}
A truncated series expansion in powers of $e$ of this function\footnote{The use of the distance squared avoids the square root, which is very convenient both for algebraic manipulation and for numerical analysis.} is
\begin{equation}
    \textrm{d}^{2(N)}_*(a,e,\alpha,\beta) = \sum_{i=0}^{N} d_i(a,\alpha,\beta)\, e^i, \label{eq:2-3-10}
\end{equation}
in which the $d_i(a,\alpha,\beta)$ coefficients need to be specified. In this contribution, the $d_i(a,\alpha,\beta)$'s have been computed for $N$ up to 6 using the {\sc Maple} 2017.3 math software:
\begin{eqnarray}
  d_i & = & 0\ \text{if }i \text{ is odd,} \label{eq:2-3-x}\\
  d_0 & = & \left( \sqrt {{\alpha}^{2}+{\beta}^{2}}-a \right)^{2}, \\
  d_2 & = & {\frac   {a{\beta}^{2} \left( \sqrt {{\alpha}^{2}+{\beta}^{2}}-a \right)}{{\alpha}^{2}+{\beta}^{2}}}, \\
  d_4 & = & {\frac {{a\beta}^{2}}{4 \left( {\alpha}^{2}+{\beta}^{2} \right) ^{5/2}}}
	\left( 4\,a {\alpha}^{2}\sqrt {{\alpha}^{2}+{\beta}^{2}} - 4\,{a}^{2}{\alpha}^{2}+ {\alpha}^{2}{\beta}^{2}+{\beta}^
{4} \right), \\
  d_6 & = & \frac{a^2 \alpha^2 \beta^{2}
  \left[ {\alpha}^{2}\left( {\beta}^{2} -{a}^{2} \right) +{a}^{2}{\beta}^{2}+{\beta}^{4} \right]}
  {\left( {\alpha}^{2}+{\beta}^{2} \right) ^{4}}
  - \frac{3 a \beta^{2} \left[ -\frac 2 3 {a}^{2}{\alpha}^{4}+ {\alpha}^{2}\left( {a}^{2}{\beta}^{2}-
\frac {{\beta}^{4}}{12} \right) -\frac {{\beta}^{6}}{12}
 \right]}{ 2 \left( {\alpha}^{2}+{\beta}^{2} \right) ^{7/2}}. \label{eq:2-3-y}
 \end{eqnarray}
The distance approximation can then be computed by substituting the required number of terms up $N=0,2,4,6$ in Eq.~\ref{eq:2-3-10}.
Again, the lower the order $N$, the more mathematically treatable the final expression (Eq.~\ref{eq:2-3-9}) and the lower the accuracy obtained for the distance.

\section{Accuracy assessment}
\label{sec:accuracy}

The original SDG-MOID method introduces negligible errors in the distance computation, i.e., it does not increase the uncertainty of the original orbital data. For an approximate MOID, like that provided by the methods presented here, the accuracy should not be lower than the minimum uncertainty of the data, the best performance in terms of computing speed being achieved when the former is close to the latter.

The uncertainty of the state vector is usually expressed by its six-dimensional confidence ellipsoid. However, since only the position accuracy is relevant here, we limit our attention to the corresponding three-dimensional confidence (or uncertainty) ellipsoid. Its major axis is aligned very closely with the velocity vector (its direction is tangent to the trajectory), whereas the other two axes define the plane normal to the orbit.

The proposed methods have been tested on the computation of the Earth MOIDs of Near Earth Objects (NEOs). In this case, Earth is the primary object and its orbital eccentricity constitutes the small parameter. Since the Earth's position vector is known with the highest possible precision \citep{Gronchi:2007b}, the uncertainty of the MOID is entirely determined by the uncertainty of the orbit of the NEO, expressed by its covariance matrix \citep{Milani:1989}.

The space objects of interest are tracked by dedicated observatories, and their data are systematically computed and recorded in catalogs. Each entry has an associated uncertainty (affecting both position and velocity) due to several causes, such as instrument precision, shortage of measures, truncations, round-off errors \citep{Gronchi:2007b}.
The orbit determination of NEOs is carried out by either radar tracking or optical observations. The former technique is characterized by low measurement errors, but it is applicable to a limited number of objects. The majority of NEOs are monitored by optical observations, which are affected by larger errors owing to the more limited visibility conditions.
The NEOs whose orbits are determined by radio telescopes are associated with the smallest position uncertainties, at the level of $100$ m, whereas optical determinations are affected by much larger (by at least two orders of magnitude) position uncertainties.

The minimum accuracy required by the approximate algorithm must be greater than the smallest uncertainty in the position vector.
Therefore, the smallest error in length (maximum tolerance) is the size of the shortest semiaxis of the $1\sigma$ position's confidence ellipsoids obtained for the subset of NEOs observed with radar tracking. These values have been calculated from the covariance matrix data in equinoctial elements of the first 281 radar-tracked Near Earth Asteroids (NEAs)\footnote{NEAs have perihelion below 1.3 au and
aphelion above 0.983 au. Some NEAs are classified as Potentially Hazardous Asteroids (PHAs), i.e., asteroids whose MOID is lower than 0.05 au and whose absolute magnitude is 22 or brighter \citep[see also][]{Novakovic:2013}. PHAs are associated with a higher probability of experiencing a close encounter with Earth.} of the NEODyS catalogue using the algorithm illustrated in Appendix \ref{app:1}. The results are the lengths of the three semiaxes of each $1\sigma$ position's confidence ellipsoid. On the other hand, the geometric mean of the two smallest semiaxes is the radius of a circle whose area is equal to that of the section (ellipse) of the ellipsoid with the plane containing the two shortest axes. This holds because the  minimum distance line is in this plane but its exact orientation is not a priori known. Note that the longest axis --corresponding to the maximum uncertainty due to instability in the sense of Lyapunov-- is tangent to the orbit, whereas the MOID always corresponds to a line that is normal to both orbits.

Table~\ref{tab:3-1} is a statistical summary of the data. The several rows contain the minimum, the average and the maximum values of the sets formed by the three semiaxis lengths (i.e., shortest, intermediate and longest) of the position's confidence ellipsoids and by the geometric mean of the shortest and intermediate semiaxes.
\begin{table}
\caption{\label{tab:3-1} Summary of the statistical properties of the $1\sigma$ position's confidence ellipsoids of the first 281 radar-tracked NEAs published in the NEODyS database.}
\centering 
\begin{tabular}{lllll}
\hline \hline
      & \multicolumn{3}{c}{Semiaxis length (km)} & Geometric mean of \\
        \cline{2-4}
         &Shortest   &Intermediate   &Longest & shortest and intermediate (km)  \\
         \cline{2-5}
  Minimum	& $1.004\times 10^{-1}$ & $4.307\times 10^{-1}$	& $7.317\times 10^{1}$	& $2.641\times 10^{-1}$ \\
  Average	& $1.839\times 10^{1}$	& $6.810\times 10^{1}$	& $6.429\times 10^{3}$	& $3.312\times 10^{1}$ \\
  Maximum	& $1.939\times 10^{2}$	& $8.055\times 10^{2}$	& $1.731\times 10^{5}$	& $3.952\times 10^{2}$ \\
  \hline
\end{tabular}
\end{table}
The minimum of the shortest is $6.711 \times 10^{-10}$ au ($\sim$ 100 m). Therefore, $\Delta l = 6.711 \times 10^{-10}$ au is the tolerance to be used to assess the level of trueness for the approximate computation of the Earth MOIDs of all the objects of the NEODyS database:
\begin{equation}
    \Delta \textrm{MOID} = | \textrm{MOID}_{true} - \textrm{MOID}_{approx.} | \le \Delta l.
\end{equation}

\section{Tests}
\label{sec:tests}

\subsection{Approximate computation of the in-plane distance for low-eccentricity ellipses}
The two families of procedures here proposed for the computation of the distance between a low-eccentricity ellipse and a coplanar point, i.e., $\textrm{d}(u_*^{(N)})$ (Eq.~\ref{eq:2-2-13}) and
$\sqrt{ {\textrm{d}}_*^{2(N)} }$ (Eq.~\ref{eq:2-3-10}), respectively, have been coded and tested.
Simulations have been run on an Intel$^{\textregistered}$ Core$^{\texttrademark}$ I7-8750H $@$2.2 GHz processor with 32 MiB of RAM.
Accuracy and computing speed tests have been executed on an ellipse with unit semimajor axis ($a$=1) and eccentricity close to that of Earth's orbit ($e$=$1.671022 \times 10^{-2}$) and a set of 616 points evenly distributed in the first quadrant and belonging to a family of concentric circles with radii respectively larger, equal and smaller than the semimajor axis of the ellipse. The center of the circles coincides with the center of the ellipse. The points form a grid with polar coordinates $(\rho,\theta) : \rho=\{2^{i-5},\ i=0,1,\dots,10\},\theta=\{\frac{j\pi}{110}, j=0,1,\dots,55\}$.

\subsubsection{Numerical errors}
Table~\ref{tab:4-1} summarizes the basic statistical properties, i.e., the minimum, the average and the maximum value, of the absolute error of the in-plane distances from the selected elliptical orbit to each of the 616 test points (see text). 
The in-plane distances have been obtained with the two families of approximate algorithms at different expansion orders. 
We observe that the $\textrm{d}(u_*^{(N)})$ algorithm is far more accurate than $\sqrt{ {\textrm{d}}_*^{2(N)}}$ at the same $N$. With only two terms, it introduces no appreciable error; the $\sqrt{ {\textrm{d}}_*^{2(N)}}$ algorithm requires four terms to reach a comparable performance.

\begin{table}
\caption{\label{tab:4-1}
Basic statistical properties  (minimum, average and maximum) of the in-plane distance error $\epsilon$ of the two approximate algorithms, i.e., $\sqrt{\textrm{d}_*^{2(N)}}$ and $\textrm{d}(u_*^{(N)})$, at different expansion orders. Tests are executed on the elliptical orbit and the grid of points defined in the text. Comparisons are made with the true distance $\textrm{d}$. }
\centering
\begin{tabular}{lllllllll}
  \hline \hline
& \multicolumn{8}{c}{In-plane distance errors (au)}	\\
\cline{2-9}
            & \multicolumn{4}{c}{$\sqrt{d_*^{2(N)}}$}       & \multicolumn{4}{c}{$d(u_*^{(N)})$} \\
\cline{2-5} \cline{6-9}
$N$         &0          &2          &4          &6          &0          &2          &4        &6\\
\hline
$\epsilon_{min}$   &  0.00 &  0.00 &  0.00 &  0.00 &  0.00 &  0.00 &  0.00 &  0.00 \\
$\epsilon_{ave}$	  &$6.98\times 10^{-5}$ & $6.38\times 10^{-6}$  & $8.39\times 10^{-10}$ & $4.90\times 10^{-13}$
            &$1.75\times 10^{-6}$ & $3.78\times 10^{-13}$ & $2.49\times 10^{-16}$ & $2.23\times 10^{-16}$\\
$\epsilon_{max}$	  &$1.40\times 10^{-4}$ & $1.40\times 10^{-4}$  & $1.30\times 10^{-8}$ & $7.61\times 10^{-12}$
            &$3.49\times 10^{-5}$ & $6.26\times 10^{-12}$ & $3.55\times 10^{-15}$ & $3.55\times 10^{-15}$ \\
\hline
\end{tabular}
\end{table}

\subsubsection{Execution speed}
Table~\ref{tab:4-2} reports results and statistics regarding the execution times for the test cases described above. 
We observe that
\begin{itemize}
  \item the two approximate methods are at least ten times faster than the exact distance determination;
  \item there are no appreciable differences in execution time between the two approximate algorithms, neither by family nor by number of terms;
  \item the computing speed is inversely proportional to the expansion order.
\end{itemize}

\begin{table}
\caption{\label{tab:4-2}
Execution times per in-plane distance determination as obtained with the approximate algorithms $\sqrt{\textrm{d}_*^{2(N)}}$ and $\textrm{d}(u_*^{(N)})$ at different expansion orders (columns 2-9) and the computing time required by the exact distance determination $\textrm{d}$ (column 10). Tests are executed on the elliptical orbit and the grid of points defined in the text. The two rows before the last provide the statistics of 10 samples obtained from a set of 12 elements after discarding the largest and the smallest. The last row includes the ratio of each mean to the highest.}
\centering
\begin{tabular}{lrrrrrrrrr}\hline \hline
& \multicolumn{9}{c}{Execution times per in-plane distance determination (ns)}	\\
\cline{2-10}
            & \multicolumn{4}{c}{$\sqrt{d_*^{2(N)}}$}       & \multicolumn{4}{c}{$d(u_*^{(N)})$} & \multicolumn{1}{c}{$d$}\\
$N$         &0          &2          &4          &6 & 0 & 2 & 4 & 6 & $(\infty)$\\ \hline
\#1			&	10.804 &	10.788	&	10.893	&	11.498	
            &	10.900 &	10.966	&	10.943	&	10.933	&	121.82	\\
\#2			&	10.938 &	10.981	&	10.959	&	10.962	
            &	10.954 &	10.932	&	10.913	&	11.086	&	122.60	\\
\#3			&	10.970 &	10.875	&	10.806	&	10.908	
            &	10.991 &	10.905	&	10.947	&	12.120	&	123.66	\\
\#4			&	10.898 &	10.798	&	10.901	&	11.077	
            &	10.970 &	10.829	&	10.796	&	11.584	&	122.95	\\
\#5			&	11.011 &	10.766	&	11.328	&	11.170
            &	10.801 &	10.968	&	10.933	&	11.757	&	121.53	\\
\#6			&	10.981 &	10.786	&	11.275	&	10.955	
            &	10.804 &	10.952	&	10.812	&	10.917	&	124.17	\\
\#7			&	10.867 &    11.213	&	10.789	& 	11.766
            & 	10.868 &	10.887	&	10.808	&	10.948	& 	122.75	\\
\#8			&	10.956 &	10.940	&	10.947	&	10.916	
            &	10.798 &	10.834	&	10.815	&	10.795	&	121.71	\\
\#9			&	10.844 &	10.994	&	10.934	&	11.499
            &	10.810 &	10.784	&	11.509	&	10.904	&	121.47	\\
\#10	    &	10.783 &	11.551	&	11.264	&	11.660
            &	10.833 &	10.707	&	10.987	&	11.225	&	122.96	\\
\hline
Mean		&	10.905 &	10.969	&	11.010	&	11.241	
            &	10.873 &	10.876	&	10.946	&	11.227	&	122.56	\\
Std. dev.	&	0.078  &	0.246	&	0.201	&	0.332
            &	0.076  &	0.087	&	0.210	&	0.444	&	0.92	\\
\hline
Mean &	8.90 &	8.95	&	8.98	&	9.17
            &	8.87	&	8.87	&	8.93	&	9.16	&	100.0\\
ratio (\%)	& & & & & & & & &\\ \hline
\end{tabular}
\end{table}

\subsection{Approximate computation of the MOID for low-eccentricity ellipses}
The $d(u_*^{(N)})$ algorithms exhibit similar execution times but better accuracy than the $\sqrt{\textrm{d}_*^{2(N)}}$ methods. For this reason, the former one-parameter family, with the expansion order as the parameter, is the method of choice to replace the exact in-plane distance computation method in SDG-MOID. Tests have been performed to assess accuracy and execution speed of this method relative to the exact SGD-MOID determination. The simulations consist in the computation of the Earth MOID of objects belonging to the NEODyS database.

\subsubsection{Numerical errors for the approximate SDG-MOID}
Table~\ref{tab:4-3} reports the most significant statistical properties of the results of the accuracy tests, i.e., the minimum, the average and the maximum absolute error of the MOID computed using the $d(u_*^{(N)})$ algorithm at different expansion orders.
The high accuracy (trueness) obtained with $N$ = 2 is the most remarkable outcome.
\begin{table}
\caption{\label{tab:4-3}
Basic statistical properties (minimum, average and maximum) of the Earth MOID error $\epsilon$  resulting from application of the $\textrm{d}(u_*^{(N)})$ family of approximate algorithms at different expansion orders relative to the exact Earth SDG-MOID for objects in the NEODyS database.
}
\centering
\begin{tabular}{lllll}
  \hline \hline
& \multicolumn{4}{c}{MOID error (au)}	\\
\cline{2-5}
            & \multicolumn{4}{c}{$d(u_*^{(N)})$} \\
\cline{2-5}
$N$         &0          &2          &4        &6\\
\hline
$\epsilon_{min}$   &0.00 &0.00 &0.00 &0.00\\
$\epsilon_{ave}$	  &$2.884 \times 10^{-7}$ &$1.103 \times 10^{-14}$ &$1.763 \times 10^{-16}$ &$1.729 \times 10^{-16}$\\
$\epsilon_{max}$	  &$6.941 \times 10^{-5}$ &$6.375 \times 10^{-11}$ &$1.195 \times 10^{-12}$ &$1.195 \times 10^{-12}$\\
\hline
\end{tabular}
\end{table}

\subsubsection{Execution speed for the approximate SDG-MOID}
Table~\ref{tab:4-4} shows the average execution times per distance computation resulting from several determinations --both exact and approximate-- of Earth MOID of NEODyS objects, as well as some descriptive statistics and relevant ratios.
On average, the $d(u_*^{(N)})$ algorithm with $N=2$ is 40\% faster than the exact SDG-MOID. This result constitutes a remarkable enhancement.
\begin{table}
\caption{\label{tab:4-4}
Average execution times per distance determination of the complete NEODyS database as obtained with the exact SDG-MOID (first column) and with the $d(u_*^{(N)})$ algorithm at $N$ = 6,4,2,0 (remaining columns). The two rows before the last provide the statistics of the 20 execution samples. The last row reports the ratio of each mean to the highest.}
  \centering
\begin{tabular}{lrrrrr}
  \hline \hline
  & \multicolumn{5}{c}{Average execution times (ns)}\\
 \cline{2-6}
  & d & \multicolumn{4}{c}{$d(u_*^{(N)})$} \\
  $N$ & $\infty$ & 6 & 4 & 2 & 0 \\
   \hline
                    & 13294	& 9846	& 8753	& 8197	& 7861 \\
                    & 13303	& 9610	& 8705	& 7981	& 7849 \\
                    & 13141	& 10455	& 8772	& 7914	& 7841 \\
                    & 13274	& 9517	& 8941	& 7847	& 7917 \\
                    & 13029	& 9912	& 8927	& 7812	& 7843 \\
                    & 13406	& 9713	& 8676	& 7781	& 7739 \\
                    & 13309	& 9554	& 8714	& 7792	& 8278 \\
                    & 13219	& 10465	& 8664	& 7933	& 8145 \\
                    & 13039	& 10167	& 8763	& 8004	& 7853 \\
                    & 13049	& 9719	& 8687	& 7931	& 7791 \\
                    & 13508 & 9846	& 8899	& 7874	& 8006 \\
                    & 13196	& 9670	& 8885	& 7900	& 7885 \\
                    & 13035	& 9581	& 8735	& 7789	& 7757 \\
                    & 13035	& 10511	& 8685	& 7802	& 7758 \\
                    & 13463	& 9580	& 8768	& 8414	& 7837 \\
                    & 14053	& 9520	& 8653	& 8288	& 7897 \\
                    & 13194	& 10042	& 8881	& 7965	& 7784 \\                    & 13048	& 9774	& 8945	& 7813	& 7755 \\
                    & 13104	& 9599	& 9008	& 7815	& 8000 \\
                    & 13535	& 10421	& 8680	& 7781	& 7820 \\
                    \hline
Mean                & 13262	& 9875	& 8787	& 7932	& 7881 \\
Std. dev.           & 249   & 347	& 113	& 177	& 136  \\
\hline
Mean      & 100	& 74.5	& 66.3	& 59.8	& 59.4 \\
ratio (\%) & & & & & \\
\hline
\end{tabular}
\end{table}

\section{Application: time monitoring of the MOID}
\label{sec:app}
\begin{figure}
\resizebox{\hsize}{!}{\includegraphics{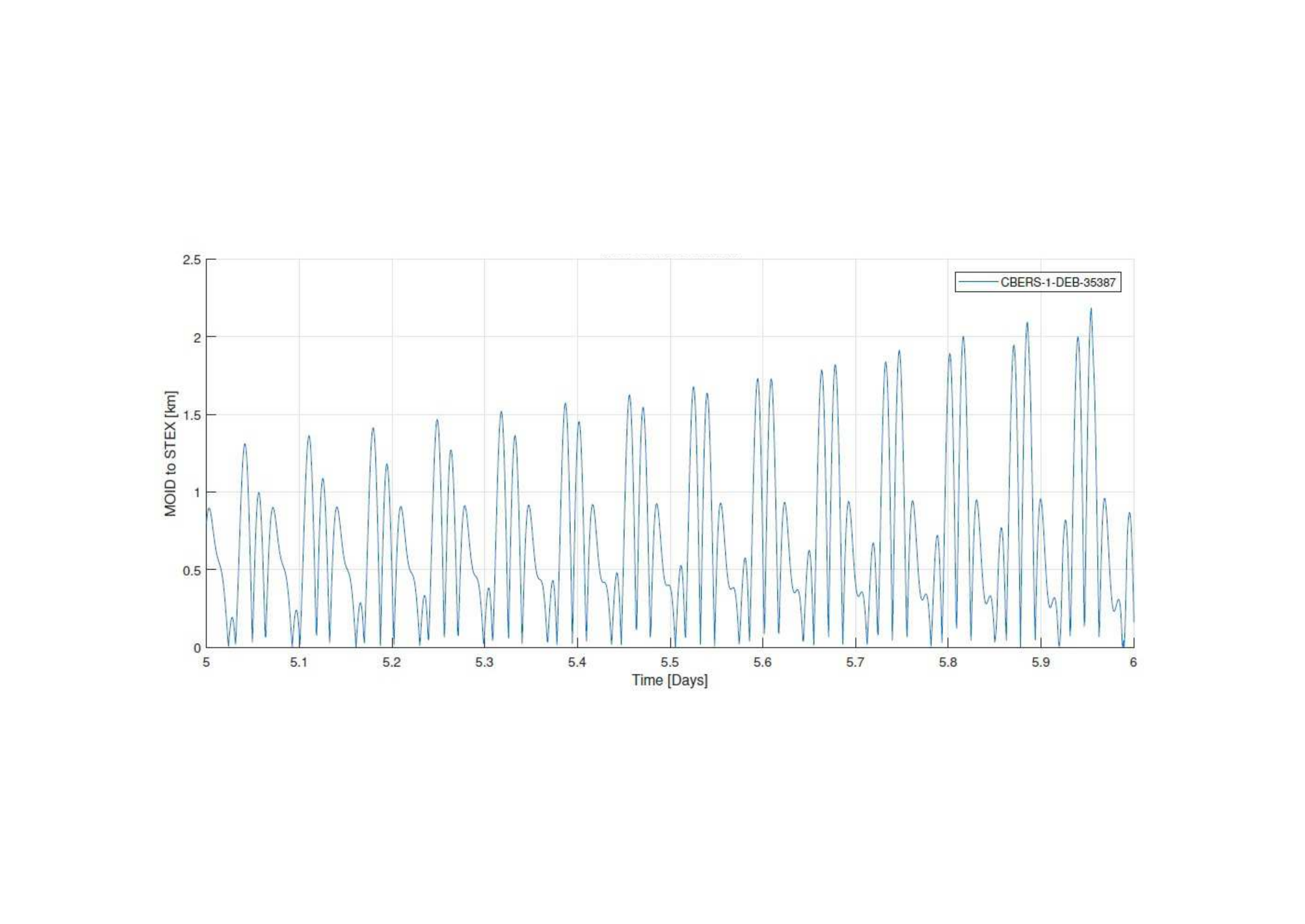}}
\caption{MOID between STEX and CBERS~1 between day 5 and day 6 after 12 pm of June $16^{th}$ 2019 (58 650.5 MJD). Simulations made with DROMOID.}
\label{fig:531}
\end{figure}

The first application of the algorithms described in this paper is within a numerical tool called DROMOID \citep{Gorno:2019} that monitors the MOID of two LEO objects (e.g., a satellite and a space debris object) over time. DROMOID implements the DROMO orbital propagator \citep{Urrutxua:2016}\footnote{The preliminary version of the tool implements only the perturbations associated with the $J_2$ coefficient of the terrestrial gravity field, but its modular structure allows to introduce any other perturbing acceleration in an easy way.}
 and the SDG-MOID algorithm.  The evolution of the two orbits over a selected time period and at a given sampling resolution is computed using DROMO. Then, the osculating classical orbital elements of the two objects are obtained at each sampling time and used to determine the evolution of the MOID. One of the advantages of the tool is that it accepts in input the space debris ephemerides of NASA's Two-Line-Element (TLE) database, which are always accessible and are considered the most reliable public source of orbital data. 
The time performance of DROMOID depends on the time resolution. However, the 2-weeks propagation of 19\,664 valid TLE ephemerides at a sampling resolution of 1 hour takes only 9.5 hours on an I5 processor.
DROMOID can efficiently predict close encounters, as shown by the following example, for the detailed analysis of which the reader is referred to \citet{Gorno:2019}. According to SOCRATES, the Space Technology EXperiments (STEX) satellite and a debris from the CBERS~1 satellite had a close approach with a minimum range of 0.638 km on June $21^{st}$ 2019 at 18:57:58.129 UTC. The two orbits have been propagated with DROMOID over a period of 7 days starting at noon of June $16^{th}$ 2019 (58 650.5 MJD) and using a sampling interval of 1 minute. The resulting MOID is smaller than 3~km during the whole propagation period and it oscillates very rapidly between 0 and 3~km, the amplitude of the variations getting smaller (below 2~km) approximately between day 2 and day 6 and the oscillations reaching the highest frequency between day 5 and day 6  (see Fig.~\ref{fig:531}). When the MOID is zero, the two orbits are crossing each other, but this not necessarily corresponds to a collision  because the latter requires time coincidence\footnote{According to SOCRATES, the collision would have occurred at a relative velocity of about 9.7 km/s, leading to the formation of a cloud of hundreds or even thousands of new space debris.}. 
Hence, SDG-MOID can be reliably employed as part of the first processing stage of a more complex tool for the real-time monitoring of thousands of space debris in the determination of the risk of collision with a spacecraft.

\section{Conclusions}
\label{sec:conclu}
Two families of algorithms for the approximate computation of the distance from an ellipse to a coplanar point have been proposed for the case
of low-eccentricity ellipses: they consist, respectively, in asymptotic expansions of
\begin{enumerate}
  \item the critical eccentric anomaly and
  \item the squared distance
\end{enumerate}
in terms of eccentricity.
The family based on the critical eccentric anomaly (Sect.~\ref{ss:2.2}, Eq.~\ref{eq:2-2-13}) is more accurate (at the same expansion order). Since the speeds of execution are very similar for the two families, this method has been chosen for further tests involving the computation of the MOID. The algorithm has been implemented in the SDG-MOID software and tested against the exact Earth MOID of NEOs. When $N=2$, the absolute Earth MOID error is always smaller than $6.4 \times 10^{-11}$ au = 9.53 m, i.e., ten times smaller than the assumed tolerance. The speed improvement over the exact Earth MOID is 40\%. 

This remarkable result makes the method based on the critical eccentric anomaly the ideal choice for all the applications in which a fast and accurate MOID computation is required: for the processing of large catalogs of objects, for the long-term monitoring of the MOID between two objects in the presence of perturbations, for the assessment of the impact probability through Monte Carlo simulations and for the propagation of the uncertainty of the observational data. 

\begin{acknowledgements}
The work of J.~M. Hedo, E. Fantino and J. Pel\'aez has been funded by Khalifa University of Science and Technology's internal grants FSU-2018-07 and CIRA-2018-85. Additionally, the authors acknowledge the support provided by the project entitled Dynamical Analysis of Complex Interplanetary Missions with reference ESP2017-87271-P, sponsored by Spanish Agencia Estatal de Investigaci{\'o}n (AEI) of Ministerio de Econom{\'i}a, Industria y Competitividad (MINECO), and by European Fund of Regional Development (FEDER).
\end{acknowledgements}

\bibliographystyle{aa}
\bibliography{asymMOID_aa}

\begin{appendix}
\section{Estimation of the positional uncertainty}
\label{app:1}

\subsection{Keplerian and Equinoctial elements}
Let $S_0=\{F;\hat{\bf i}_0, \hat{\bf j}_0, \hat{\bf k}_0\}$ be the Barycentric Celestial Reference Frame (BCRF). Let
$k$ be the index associated with an object orbiting the Sun and $a_k, e_k, \Omega_k, i_k, \omega_k, M_k$ its Keplerian orbital elements  with respect to the BCRF (Table~\ref{tab:A-1}).
The size and shape of the orbit are defined by the semi-major axis $a_k$ and the eccentricity $e_k$, respectively, whereas the elements of the set $\{\Omega_k, i_k, \omega_k\}$  are the three Euler angles defining the orientation of the perifocal reference frame $S_k=\{F;\hat{\bf u}_k, \hat{\bf v}_k, \hat{\bf w}_k\}$ in the BCRF. $M_k$ is the mean anomaly that serves to position the object on the orbit at the epoch (temporary data).
The basis $\{\hat{\bf u}_k, \hat{\bf v}_k, \hat{\bf w}_k\}$ of the perifocal reference frame as well as the rotation matrix $\left[{Q_{k0}}\right]$ (which is formed by the components of these unit vectors) are functions of the elements of the triple \{$\Omega_k, i_k, \omega_k$\} and they can be easily computed.

Let ${\bf N}_{k0}^a = \displaystyle \hat{\bf k}_0 \times \hat{\bf w}_k$ be the vector to the ascending node of object $k$ relative to the $Fx_0y_0$ plane in the BCRF. The angles in Table~\ref{tab:A-1} are defined as
\begin{eqnarray}
\Omega_k &= &\arctan\!2({\bf N}_{k0}^a\cdot \hat{\bf j}_0,{\bf N}_{k0}^a\cdot \hat{\bf i}_0), \label{eq:RAAN}\\
i_k      &= &\arccos(\hat{\bf k}_0 \cdot \hat{\bf w}_k), \\
\omega_k &= &\arctan\!2({\bf N}_{k0}^a\cdot  \hat{\bf v}_k, {\bf N}_{k0}^a \cdot \hat{\bf u}_k),\\
M_k      &= &\sqrt{\frac{\mu}{a_k^3}}\,(t_0-\tau_k),
\end{eqnarray}
where $t_0$ is the epoch, $\tau_k$ is the epoch of pericenter passage, $\mu$ is the gravitational parameter of the Sun and $\arctan\!2(y,x)$ is the angle measured in counterclockwise direction from $\hat{\bf i}$ to $x\hat{\bf i} + y\hat{\bf j}$ and defined in the interval $[0, 2\pi[$.
\begin{table}
  \caption{\label{tab:A-1} The Keplerian orbital elements}
  \centering
  \begin{tabular}{lll}
    \hline \hline
    Name             & Notation & Range      \\ \hline
    Semi-major axis        & $a$            & $\in [0,\infty]$ \\
    Eccentricity           & $e$            & $\in [0,\infty]$ \\
    Right ascension        & $\Omega$       & $\in [0,2\pi]$   \\
    Inclination            & $i$            & $\in [0,\pi]$    \\
    Argument of pericenter & $\omega$       & $\in [0,2\pi]$   \\
    Mean anomaly at epoch  & $M$            & $\in [0,2\pi]$   \\ \hline
  \end{tabular}
\end{table}

When $\sin i_k=0$, $\Omega_k$ is undefined because so is the nodal line. When $e_k=0$, $\omega_k$ is undefined because so is the pericenter. The singularities that arise in many applications, e.g., orbit determination and orbit propagation, under these circumstances are well known \citep{Broucke:1970}. Given the high frequency of occurrence of low-inclination and/or low-eccentricity orbits in the Solar System, an alternative set of elements are often employed with respect to which the above equations and expressions are non-singular. It is the set of so-called equinoctial elements $a_i \ (i=1,2,\dots,6)$ \citep{Arsenault:1970} (see Table~\ref{tab:A-2}). Every orbit can be associated to an equinoctial reference frame $S'_k = \{F; \hat{\bf f}_k, \hat{\bf g}_k, \hat{\bf w}_k\}$, where $\hat{\bf f}_k$ and $\hat{\bf g}_k$ define the orbital plane, and the angle between $\hat{\bf f}_k$ and ${\bf N}_{k0}^a$ is the longitude of ascending node. The first and the sixth equinoctial elements are the semimajor axis and the mean longitude, respectively. The second and the third are the components of the eccentricity vector (${\bf e}_k = a_2 \hat{\bf f}_k + a_3 \hat{\bf g}_k$) in the equinoctial reference frame, whereas the fourth and the fifth are the components of the ascending node vector ($\tan(\frac i 2){\bf n}_{k0}^a = a_4 \hat{\bf f}_k + a_5 \hat{\bf g}_k$).
\begin{table}
  \caption{\label{tab:A-2} The equinoctial elements (tensor and two common notations N$_1$ and N$_2$) and their relation with the Keplerian elements.}
  \centering
  \begin{tabular}{llll}
  \hline
   \multicolumn{3}{c}{Notation}   &        \multicolumn{1}{c}{Relation with} \\ \cline{1-3}
   Tensor &  N$_1$ & N$_2$ &  \multicolumn{1}{c}{Keplerian elements} \\ \hline
  $a_1$ & $a$   & $a$ & $a$ \\
  $a_2$ & $P_1$ & $h$ & $e \sin{(\Omega+\omega)}$ \\
  $a_3$ & $P_2$ & $k$ & $e \cos{(\Omega+\omega)}$ \\
  $a_4$ & $Q_1$ & $p$ & $\tan{(\frac i 2)}\sin\Omega$ \\
  $a_5$ & $Q_2$ & $q$ & $\tan{(\frac i 2)}\cos\Omega$ \\
  $a_6$ & $l$   & $\lambda$ & $\Omega+\omega+M$ \\
  \hline
\end{tabular}\\
\end{table}

\subsection{Positional uncertainty from the equinoctial covariance matrix}
In the following, we present the positional uncertainty estimation of an object given its set of equinoctial elements and the corresponding covariance matrix. The velocity uncertainty estimation must be computed as well, but it is not relevant for the MOID problem.
Knowing the positional uncertainties of the two objects allows to select the appropriate accuracy in their MOID computation. As a matter of fact, requiring a higher accuracy than that of the input data slows down the computations without any benefit.

Let ${\bf r}, \dot{\bf r}$ be the position and velocity vectors of the center of mass of a given object (here the index $k$ has been omitted for the sake of brevity) in a quasi-inertial frame of reference --for example, the BCRF for objects orbiting the Sun.
The former vectors can be expressed as linear combinations of basis unit vectors $\{\hat{\bf u}_1,\hat{\bf u}_2,\hat{\bf u}_3\}$ of the reference frame
\begin{alignat}{5}
{\bf r} &= x_i \hat{\bf u}_i &;\dot{\bf r} &= \dot{x}_i {\bf u}_i \quad(i=1,2,3). \label{eq:1}
\end{alignat}
Here, we adopt Einstein summation convention and the dot symbol above a variable represents time differentiation. When the latter is applied to a vector, it is done in a quasi-inertial reference frame. The state vector ${\bf R}$ is defined as
\begin{equation}
{\bf R} = \begin{pmatrix} x_1 & x_2 & x_3 & \dot{x}_1 & \dot{x}_2 & \dot{x}_3  \end{pmatrix}^T.
\end{equation}
The position and velocity vectors can be written as functions of the equinoctial elements $a_i$  ($i=1,2,\dots,6$) \citep{Danielson:1995}:
\begin{alignat}{2}
  {\bf r} = {\bf r}(a_j) \;\; \dot{\bf r} &= \dot{\bf r}(a_j) \quad (j=1,2,\dots,6). \label{eq:3}
\end{alignat}
Introducing the vector ${\bf a}=\begin{pmatrix} a_1 & a_2 & \dots & a_6 \end{pmatrix}^T$ of equinoctial elements allows to express Eq.~\ref{eq:3} as a vector-valued function of a vector variable:
\[
{\bf R} = {\bf R}({\bf a}).   \tag*{(\ref{eq:3}')}
\]
Differentiation yields
\begin{equation}
  \Delta{\bf R} = {\bf J} ({\bf a}) \Delta{\bf a}, \label{eq:4}
\end{equation}
where ${\bf J} \in\mathbb{M}_{6\times 6}$ is the Jacobian matrix of ${\bf R}({\bf a}): J_{ij} = \frac{\partial R_i}{\partial a_j}$. 
Solving Eq.~\ref{eq:4} with respect to $\Delta {\bf a}$ (which is possible if and only if ${\bf J}$ is regular) provides
\begin{equation}
  \Delta{\bf a} = {\bf J}^{-1}({\bf a})\,\Delta{\bf R}.  \label{eq:5}
\end{equation}
Under the usual assumption that ${\bf a} \sim \mathcal{N}_6({\bf \mu}_a,{\bf \Sigma}_a)$, the probability density function $f({\bf a}) $ of the equinoctial elements is \citep{Wiesel:2003}
\begin{equation}
    f({\bf a}) = (2\pi)^{-3} |{\bf \Sigma}_a|^{-1/2} \cdot \exp\left(-\frac 1 2 ({\bf a}-{\bf \mu}_a)^T {\bf \Sigma}_a^{-1} ({\bf a}-{\bf \mu}_a)\right).
\end{equation}
In order for the integral of $f$ (which gives the probability) to be convergent, $({\bf a}-{\bf \mu}_a)^T {\bf \Sigma}_a^{-1} ({\bf a}-{\bf \mu}_a)$ must be non-negative. This requires that ${\bf \Sigma}_a^{-1}$ (the precision or normal matrix) and its inverse ${\bf \Sigma}_a$ (the covariance matrix) be positive definite. The precision matrix defines a quadratic form \citep{Gronchi:2007} used to calculate the sum of the squares of the residuals $\Delta{\bf a}= {\bf a}-{\bf \mu}_a $. This quantity can be employed as the objective function of an optimization problem.
The locus of the points with identical probability is a six-dimensional ellipsoid centered in the nominal orbit (six-dimensional point). The ellipsoid corresponding to a significance level $p$ is
\begin{equation}
   \left(\frac \kappa{\sigma}\right)^2 = \Delta{\bf a}^T\,{\bf \Sigma}^{-1}({\bf a})\,\Delta{\bf a},  \label{eq:6}
\end{equation}
where $\sigma$ is the standard deviation, $\kappa$ is the radius of the probability interval (usually expressed in $\sigma$'s) and the non-dimensional parameter $\frac \kappa{\sigma}$ is the square root of the quantile $p$ of a $\chi^2$ distribution of $6$ degrees of freedom, because $\Delta{\bf a} \sim \mathcal{N}_6 ({\bf 0}, {\bf I})$ if it is properly rotated and scaled. ${\bf 0}$ is the six-dimensional null vector and ${\bf I}$ is the six-dimensional unit matrix. Introducing Eq.~\ref{eq:5} into Eq.~\ref{eq:6} yields
\begin{alignat}{2}
  \left(\frac \kappa{\sigma}\right)^2 &= \Delta{\bf R}^T \underbrace{({\bf J}^{-1})^T \,{\bf \Sigma}_a^{-1}({\bf a})\,{\bf J}^{-1}({\bf a})}_{{\bf \Sigma}_R^{-1}}\,\Delta{\bf R},
 \end{alignat}
which shows that the matrix ${\bf \Sigma}_R^{-1}$ is a tensor of rank 2. Besides, ${\bf \Sigma}_R^{-1}$ and ${\bf \Sigma}_R$ are regular, symmetric and positive definite.

Since we are interested in positional errors only, we are going to limit our attention to the $3 \times 3$ principal sub-matrix of ${\bf \Sigma}_R$ involving exclusively positional covariances (${\bf \Sigma}_{r}$). In its principal axis ($\{O_i;x'_i, y'_i, z'_i\}$), the corresponding matrix ${\bf \Sigma}_{r'}^{-1}$ is diagonal, and the ellipsoid takes the canonical form
\begin{equation}
   \left(\frac \kappa{\sigma}\right)^2 =\Delta {\bf r}^T {\bf \Sigma}_{r}^{-1} \Delta{\bf r} = \Delta{\bf r}'^T {\bf \Sigma}_{r'}^{-1} \Delta{\bf r}' = \sum_{i=1}^3 \sigma_{x'_i}^{-2} {\Delta x'_i}^2
   \Rightarrow    1 =  \sum_{i=1}^3 \left(\frac{\Delta x'_i}{\left(\frac \kappa{\sigma}\right) \sigma_{x'_i}}\right)^2.  \label{eq:3DEll}
\end{equation}
in which $\left(\frac \kappa{\sigma}\right)^2$ is the quantile $p$ of a $\chi^2$ distribution of $3$ degrees of freedom, and $l_i = \left(\frac \kappa{\sigma}\right)\sigma_{x'_i}$ is the $i^{th}$ semi-axis length.

In conclusion, by solving the spectral problem for ${\bf \Sigma}_{r}$, we obtain the eigenvalues of the covariance matrix, which are the critical variances:
\begin{alignat}{2}
  \textbf{Spectrum}({\bf \Sigma}_{r}) &= \{\sigma^2_{x'_1}, \sigma^2_{x'_3}, \sigma^2_{x'_3}\}.
\end{alignat}
Then, by adopting a conservative approach, the minimum error $\mathcal{E}_{min}$ associated with a positional uncertainty is
\begin{equation}
  \mathcal{E}_{min} = \min_{i}\ l_i = \left(\frac \kappa{\sigma}\right)\min_{i} {\sigma_{x'_i}}.
\end{equation}
where $\frac \kappa{\sigma}$ is a function of $p$. It is usually set to $1$, which implies that for a 3-D ellipsoid $p\approx 0.2$.

\end{appendix}

\end{document}